# High & ultra-high energy cosmic ray acceleration in structure-formation shocks in Abell 3376 galaxy cluster

J. Bagchi[a], Florence Durret[b], Gastao B. Lima Neto[c], Surajit Paul[a], Satyajit Chavan[d]
*(a) The Inter-University Centre for Astronomy and Astrophysics, Pune University Campus, Pune, India*
*(b) Institut d'Astrophysique de Paris, France  (c) Instituto Astronomico e Geofisico, Sao Paulo, Brazil*
*(d) Research Fellow (ISRO-Project), Dept. of Physics, S.R.T.M. University, Nanded, India*

Presenter: J. Bagchi (joydeep@iucaa.ernet.in), ind-bagchi-J-abs1-og14-poster

We describe multiwavelength optical, X-ray and radio evidences of energetic mergers in southern galaxy cluster Abell 3376 (A3376, redshift z=0.046, X-ray luminosity $L_X(0.1-2.4\text{ keV}) = 2.48 \times 10^{44}\text{ erg s}^{-1}$). Here we report the rare discovery of gigantic, Mpc-scale non-thermal radio structures ('radio-arcs'), shaped like a pair of bow-shock fronts on the merger axis. We argue for the shock acceleration origin of the radio emission and discuss the possibilty of acceleration of very high energy cosmic ray particles in this and similar clusters.

## 1. Introduction

The origin of ultra high energy cosmic ray particles (UHECR) with surprisingly large energies of $\sim 10^{18}-10^{20}$ eV is a long-standing enigma of Physics [1]. In any acceleration scheme the radiation background induced energy losses and source size/energy constraints put a limit to attainable maximum energy. Observationally, cosmic ray sources of required parameters are yet unidentified. But, some potential accelerator sources are radio galaxies (lobes, jets), pulsar magnetospheres, gamma-ray bursts, and galaxy clusters [2]. Clusters of galaxies are the largest known virialised structures which assemble by gravitational infall of smaller mass components. Accretion, mergers and collisions are the dominant processes of 'structure formation'. During cluster mergers the enormous kinetic energy of colliding subclusters ($\sim 10^{63-64}$ erg) is dissipated in the form of shock-waves which play a pivotal role in heating of the intra-cluster medium (ICM) to the virial temperatures. Large scale structure-formation shocks in the presence of magnetic fields are also known to accelerate particles upto ultra-relativistic limit by a diffusive shock acceleration process (DSA or Fermi-I) [3].

## 2. Observations of Abell 3376 in optical, X-ray & radio wavelengths

In A3376 we find several evidences of ongoing energetic structure-formation processes (subcluster mergers, IGM accretion). The optical galaxies are distributed in multiple sub-groups extending along a position angle of $\approx 70^o$, defining a probable filamentary merger axis [4, 5]. An $\approx 12$ ks exposure *ROSAT* PSPC X-ray image shown in Figure 1 further confirms the merger scenario. It reveals that the intra-cluster gas is in highly disturbed, unvirialised state; the bremsstrahlung X-ray emission elongated almost parallel to the optical axis of the merging/colliding sub-groups. We also analysed the XMM-Newton archival 47 ks observation with MOS1, MOS2 and PN camera. Figure 2 shows the thermal emission morphology (in 0.4 - 8 keV band) limited by XMM-Newton FOV to within a $20 \times 20$ arcmin$^2$ region around the peak emission. This map is quite similar to *ROSAT* iamge, but shows more clearly the 'bullet head' and multiple X-ray peaks to the south south-west of it, each one probably associated with a merging group.

The X-ray temperature map derived from XMM-Newton data is displayed in Figure 2. Typical errors on temperatures in this map are about 10%. It reveals an overall temperature of about 5 keV, with several alternate hot and cold regions crossing the cluster, divided along a prominant 'cold-arc' at about 3 keV, which originates at the north edge and curves southwards towards the east. As already argued for another merging cluster Abell 85, this strongly suggests that mergers are taking place [6], and numerical simulations show temperature maps similar to our XMM-Newton data [7].

Finally, in Figure 1 we show a deep 1.4 GHz radio map of this cluster, made from our VLA observations in its D & B/C configurations. It reveals perhaps the most interesting aspect of this cluster: a pair of very large ($\sim$ Mpc) and diffuse radio sources, located at the opposite ends of the extended X-ray emitting gas, about $36'$ or $\sim 1.9\,h_{70}^{-1}$ Mpc apart from each other. What are these giant structures ? There is no evidence for any optical galaxy obviously associated with the radio arcs, and hence they are unlikely to be the usual cluster radio galaxies. It is also implausible that they are radio-lobe pair of an active giant radio galaxy (GRG) - we detect no obvious radio jets or plumes linking them with any central optical galaxy - and GRGs of this size ($\sim 2\,h_{70}^{-1}$ Mpc) are extremely rare [8]. On the other hand, their concave bow-shock like structure, symmetric and tangential juxtaposition on the merger axis - tangential both to the chain of sub-clusters of galaxies and to the X-ray emission elongation axis - argue strongly that these arcs are part of this cluster and very possibly originate in some cosmological-scale energetic event linked to the violence of cluster merger activity. We point out the extreme rarity of occurence of such double 'radio-relics'- the only other known example of this phenomenon is a similar radio and X-ray morphology observed in Abell cluster A3667 [9].

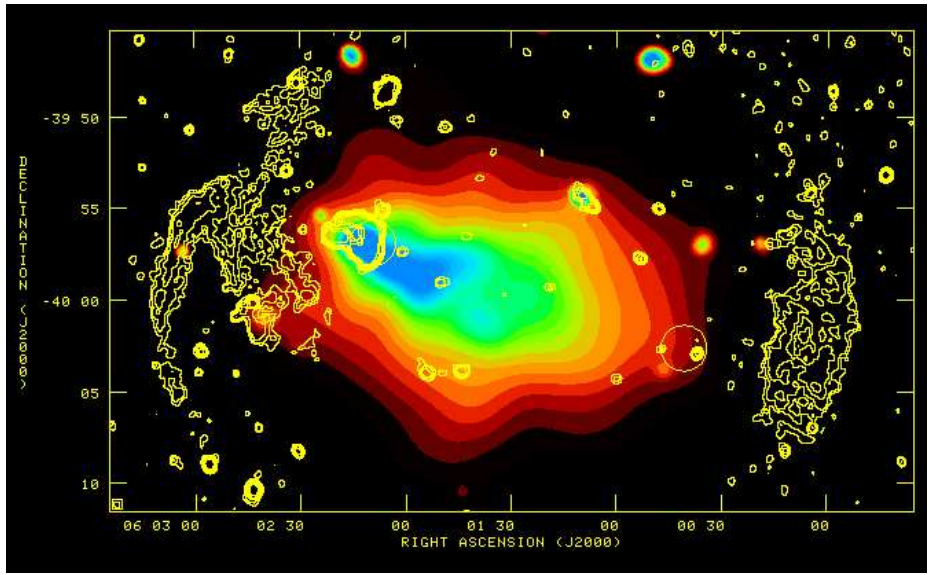

**Figure 1.** The ROSAT PSPC broad band X-ray image of A3376 cluster. Circles locate the two brightest clusters galaxies - brightest cD galaxy on lower right, second brightest (radio galaxy MRC 0600-399) near X-ray peak - and their sub-clusters [4, 5]. Contours depict deep (35 $\mu$Jy rms noise) VLA 1.4 GHz radio image at a resolution of 20 seconds of arc.

## 3. Discussion & outlook for future

The detection of these large radio structures ($L_{arc} \sim$ 1Mpc) at $\sim$ 1Mpc projected distance from cluster center requires some form of *in situ* acceleration mechanism for particles and the magnetic field powering them. This follows from comparison of radiative life time $t_{IC}$ for an electron $t_{IC} \approx 2.3 \times 10^8 \left(\frac{\gamma}{10^4}\right)^{-1} (1+z)^{-4}$ yr (assuming only inverse Compton scattering (IC) on 2.7 K background and magnetic field $B < 3$ $\mu$G), and the diffusion length within the IC cooling time: $l_{diff} \approx (D_B\, t_{IC})^{1/2} = 11.36 \left(\frac{B}{\mu G}\right)^{-1/2}$ pc $<< L_{arc}$ (here $D_B$: Bohm diffusion coefficient). For relativistic electrons $\gamma \approx 1.8 \times 10^4$ ($\sim 10$ GeV) for observed 1.4 GHz frequency and B=1 $\mu$ G. The discrepancy between the Bohm diffusion length-scale and the radio structure size is so large that, even with inclusion of advective transport by bulk flows and more effective diffusion in ordered magnetic fields, electrons are still unable to cross the emission region within a radiative life-time. Therefore, initial acceleration at a central source such as an AGN and diffusive transport upto $\sim$Mpc scale is not possible.

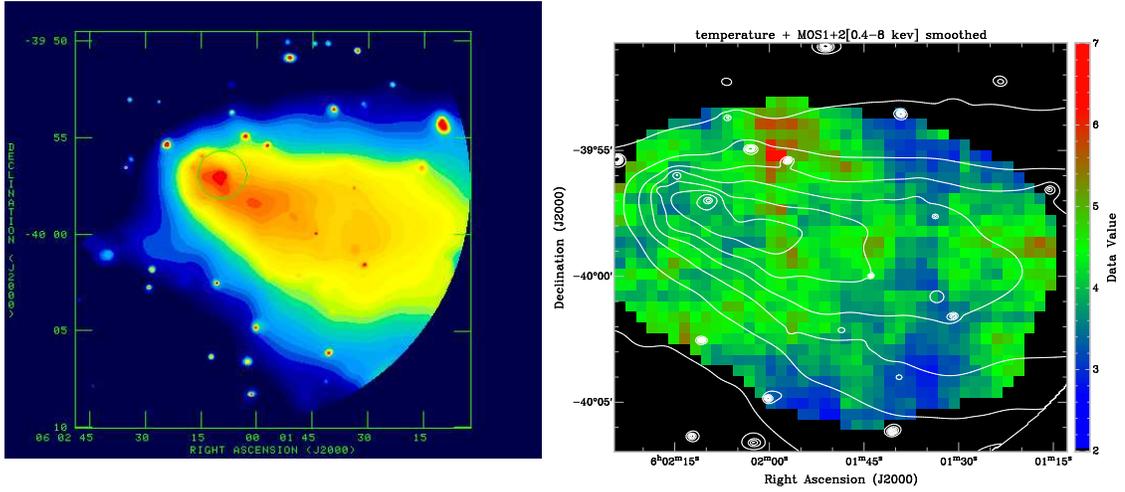

**Figure 2.** Left: XMM-Newton X-ray emission map of A3376. The circle shows the X-ray emission from the strong wide-angle tail (WAT) radio-galaxy MRC 0600-399 found at X-ray peak (see Fig. 1). Right: X-ray temperature map derived from XMM-Newton data. Temperature scale is in keV. The X-ray emission intensity iso-contours are also shown.

Further detailed multi-wavelength observations and computational study of the phenomenon are necessary in order to clarify the details of acceleration process involved. Nevertheless, the most plausible mechanism is the diffusive shock acceleration [10, 11] of cosmic ray charged particles (in general 'cosmic rays': $e^{\pm}, P^{+}, \pi^{\pm/0}, \nu, \gamma$-rays etc.) on the shock fronts coinciding with the radio structures. The morphology of these arcs is reminiscent of a classical 'bow shock' that preceeds a blunt-body in supersonic flight within a fluid. The likely origin of shocks is supersonic accretion of intergalactic matter during structure formation along a filamentary axis ('accretion shocks'). Alternatively, these structures could be shocks emanating from subcluster merger events, $\sim 1$ Gy in past from within the cluster, which led to the formation of binary shock fronts ('merger shocks') [12].

The gasdynamics induced by these processes: large-scale bulk flows, turbulence, collisionless shocks and modified magnetic fields, all create an ideal environment for cosmic ray acceleration in stochastic Fermi-I mechanism [13], and for amplification of seed magnetic fields [14]. Determination of several key shock parameters like Mach number, temperature/pressure jump structure and flow velocities await more detailed X-ray observations. However, from very general principles based on DSA framework, we can obtain some critical parameters of particle acceleration in a merging system such as A3376. The mean acceleration timescale $t_{acc}(E)$ to reach energy E is determined only by velocity jump at the shock and the diffusion coefficients [13], i.e., $t_{acc}(E) = \frac{3}{(u_1-u_2)}[(D_1/u_1) + (D_2/u_2)] \approx (8/u_1^2)\, D_B$. Here $u_1(u_2)$ is the up(down) stream flow velocity and $D_{1,2}$ are the respective diffusion coefficients. We consider a strong shock of compression ratio r=$(u_1/u_2)$=4, a frozen-in field condition: $(B/\rho) = const.$, $(D/u) = const.$, and Bohm diffusion. Under these (somewhat ideal) conditions $t_{acc} = 8.45 \times 10^5 u_3^{-2} E_{15} B_\mu^{-1} Z^{-1}$ yr, where $E_{15}=(E/10^{15}\,eV)$, $u_3=(u_1/10^3$ km s$^{-1})$, $B_\mu=(B/10^{-6}\,G)$, and $Ze$ is the ionic charge. The DSA results in a power-law particle momentum function $f(p) \propto p^{-b}$, where p is momentum and b is the power-law slope. The synchrotron/IC spectrum should also be a power-law $I(\nu) \propto \nu^{-\alpha}$, where spectral index $\alpha = (b-3)/2$. The index b and compression r are related by $b = 3r/(r-1)$, and $r = 4$, $\alpha = 0.5$ for a strong shock in a gas of specific heat ratio $\Gamma = 5/3$. Downstream of shock, the spectrum gradually steepens due to electrons undergoing diffusion and advection with the fluid flow and suffering radiative losses. For cosmic ray protons, which suffer negligible radiative losses below 50 EeV, the highest acceleration energy $E_{max}^P$ is limited by the finite life time of shocks, i.e., $t_{acc} = t_{merger} \sim 10^{9-10}$ yr, thus giving $E_{max}^P \sim 10^{18-19}$ eV. The heavier ions with $Z > 1$ (say 26Fe) can be accelerated to order of magnitude higher energies. For cosmic ray electrons the significant radiative losses limit their energy to a maximum $E_{max}^e \sim 3.73 \times 10^{13} u_3 B_\mu^{1/2}$ eV. Lower energy electrons of enegry $\sim 10$ GeV give rise to the 1.4 GHz radio emission. Hydrodynamical simulations [15] show that during gravitational in-fall and mergers in clusters, protons can be accelerated upto the GZK cut-off energy $\sim 50$ EeV, if a turbulent magnetic field of $\sim 1\,\mu G$ is available, and if about $10^{-4}$ fraction of infalling kinetic energy can be converted as high energy particles. It is likely that A3376 is a potential source which fits these conditions. The equipartition magnetic field strength in radio-arc regions is quite strong; $0.5 - 2.5 \mu G$ (depending on model parameters).

Protons will remain confined within cluster volume for time scale>> age of cluster until their energy approaches $E^P \gtrsim 2 \times 10^{17} B_\mu$ eV (Bohm diffusion). At redshift z=0.04, cluster A3376 is near enough that some of these primary or secondary cosmic ray components could actually be detected by new cosmic ray telescopes, particularly the energetic inverse Compton hard X-ray/$\gamma$-ray photons upto extreme energies $E_\gamma \sim 100\,\gamma_7^{e\,2}$ GeV (where $\gamma_7^e$ is the electron Lorentz factor in units of $10^7$). Identification of radio or hard X-ray/$\gamma$-ray emission from structure formation shocks holds a great promise for advancing current knowledge on shock formation in the IGM [12]. It should provide the first direct evidence for such shocks, revealing the underlying large-scale cosmological flows [3]. When combined with hard X-ray or $\gamma$-ray detection, the radio signal will provide a direct measure of the unknown magnetic fields in the IGM, thus providing a better understanding of processes leading to IGM magnetization and also acceleration and propagation of ultra-high energy cosmic rays.

## References

[1] Takeda M., et al., 1999, ApJ 522, 225
[2] Nagano M., Watson A.A., 2000, Rev. Mod. Phys. 72, 689
[3] J. Bagchi et al., 2002, New Astronomy, Vol. 7, 249-277
[4] Dressler A., Shectman S.A., 1988, AJ 95, 985
[5] Escalera E., et al., 1994, ApJ 423, 539
[6] Durret F., Lima Neto G.B. & Forman W., 2005, A&A 432, 809
[7] Bourdin H., Sauvageot J.-L., Slezak E., Bijaoui A. & Teyssier R., 2004, A&A 414, 429
[8] Schoenmakers A.P., de Bruyn A.G., Rottgering H.J.A., & van der Laan H., 2001, A&A 374, 861
[9] Rottgering H.J.A, Wieringa M.H., Hunstead R.W. & Ekers R.D., 1997, MNRAS 290, 577
[10] Bell A.R., 1978, MNRAS 182, 147
[11] Blandford R.D. & Ostriker J.P., 1978, ApJ Lett. 221, L29
[12] Miniati F., 2003, MNRAS 342, 1009
[13] Drury L.O'C., 1983, RPPh 46, 973
[14] Kulsrud R., Cen R., Ostriker J.P., & Ryu D., 1994, ApJ 480, 481
[15] Kang H., Ryu D. & Jones T.W., 1996, ApJ 456, 422